# Prediction of Leakage Current and Depletion Voltage in Silicon Detectors under Extra-Terrestrial Radiation Conditions

**A. Grummer[1], M.R. Hoeferkamp[1], S. Seidel[1*]**

[1] Department of Physics and Astronomy, University of New Mexico, 210 Yale Blvd. NE, Albuquerque, NM 87106 USA

**\* Correspondence:**
Sally Seidel
seidel@unm.edu

## Abstract

Silicon detection is a mature technology for registering the passage of charged particles. At the same time it continues to evolve toward increasing radiation tolerance as well as precision and adaptability. For these reasons it is likely to remain a critical element of detection systems associated with extra-terrestrial exploration. Silicon sensor leakage current and depletion voltage depend upon the integrated fluence received by the sensor, and upon its thermal history during and after the irradiation process. For minimal assumptions on shielding and hence on particle energy spectrum, and using published data on Martian ground temperature, we predict the leakage current density and the depletion voltage, as a function of time, of silicon sensors deployed continuously on the Mars surface for a duration of up to 28 Earth-years, for several sensor geometries and a worst-case temperature scenario.

## 1    Introduction

This article reports predictions of the silicon sensor bulk characteristics, leakage current density and depletion voltage, under the conditions of temperature and radiation, at the Mars surface, over a period corresponding to approximately 28.2 Earth-years, combined with the conditions of temperature and radiation associated with transit from the Earth to Mars during a period of about 250 days. The effects of the continuous flux of solar protons and galactic cosmic rays as well as occasional solar energetic particle (SEP) events are included.

Silicon detectors of moderate radiation tolerance are deployed already for numerous space applications, for example:

- Human dosimetry of charged particles is accomplished on the International Space Station/ISS through thick and thin silicon diodes [1].
- Particle radiation at the surface of Mars is measured by a suite of devices including three silicon p-i-n diodes, in the Radiation Assessment Detector (RAD) on the Mars Space Laboratory [2].
- The Cosmic Ray Telescope for the Effects of Radiation (CRaTER), an instrument on the Lunar Reconnaissance Orbiter spacecraft designed to characterize the lunar radiation environment, uses six silicon disks [3]; its records of intense SEPs are input to human organ dose projection code to assess risks to space travelers due to SEP exposure.

The primary motivation for this article is interest in the level of radiation hardness needed for a mission to Mars for the duration indicated above. Such a mission could additionally be interpreted as a proxy for extended intervals of space travel, possibly involving shielded human explorers at new frontiers, and including transport through regions susceptible to SEPs. A second motivation is presented by the resource-limited environment of such a mission (which has analogs in the environments of particle physics experiments), and a desire to investigate what parameters might permit best usage of resources. The RAD system provides examples of choices made in the resource limited environment: 4.2W are needed to power the full RAD detector, and due to power constraints it is powered for only 15 of every 60 minutes [4].

The response of silicon detectors to non-ionizing energy loss (NIEL) by through-going particles has been parameterized. The resulting Hamburg Model [5,6,7] has been widely used to predict changes in the effective dopant concentration - and consequently, in device leakage current and depletion voltage - due to radiation-induced bulk defects that evolve with temperature and time. Previous applications of this model have centered on, and been validated [8] by, experiments such as those at the CERN Large Hadron Collider. The implementation of the Hamburg Model can be found in Ref. [9].

The simulation of silicon detector leakage current and depletion voltage takes as input fluence $\Phi_{eq}$ (in 1-MeV-neutron-equivalent ($n_{eq}$) per area), temperature, and duration (which incorporates both duration of radiation exposure and of subsequent annealing). We estimate four separate contributions to $\Phi$: chronic radiation on the Mars surface; chronic radiation during transit; radiation due to SEPs that reach the Mars surface; and radiation due to SEPs that occur during the transit.

To estimate the chronic charged particle fluence at the Martian surface, due to solar protons and galactic cosmic rays, we begin with the average charged particle dose rate $D$ of (0.210±0.040) mGy/day as recorded [10] by the Curiosity Rover during the period 2012-2013.

Radiation that reaches the Martian surface includes approximately 11%-14% alpha particles and heavier nuclei, and about 1% electrons, with the remaining 85% - 90% being protons [10]. Simulations of the effect of alpha particles on silicon bulk damage indicate that their contribution to NIEL in the silicon is 0.18% that of protons [11], under the conditions of a thin layer of shielding and including the effects of secondary particles. Accordingly, we restrict the fluence considered in this simulation to that of protons. To accomplish this, in calculations below, we scale the dose rate associated with the full charged particle spectrum by 0.9. Thus we use $D_p = 0.9 \cdot 0.210$ mGy/day = 0.189 mGy/day. This dose rate is applied for a period $\tau$ = 28.2 Earth-years (15 Martian-years), which is 10293 Earth-days; this time frame is chosen because it is approximately a factor of two greater than the operational span of the Opportunity Mars Exploration Rover, which holds the record for longevity. Thus the baseline lifetime proton dose is 0.0000189 × 10293 = 1.95 Gy.

To convert this dose to fluence $\Phi$, we neglect effects associated with production of high energy electrons in the silicon, as these have no significant impact on bulk damage. With the neglect of electron production, linear energy transfer (LET) can be replaced by rate of energy loss per unit length ($dE/dx$). With this approximation,

$$\Phi \approx D_p \cdot \tau \cdot \frac{6.24 \times 10^9 \left[ \text{MeV g}^{-1} \text{ Gy}^{-1} \right]}{dE/dx}, \qquad (1)$$

where $dE/dx$ is measured in units of MeV g$^{-1}$ cm$^2$ [12,13].

The rate of energy loss per unit length ($dE/dx$) for proton kinetic energies ranging from 1 keV to 10 GeV is shown in Figure 1, using data taken from Ref. [14].

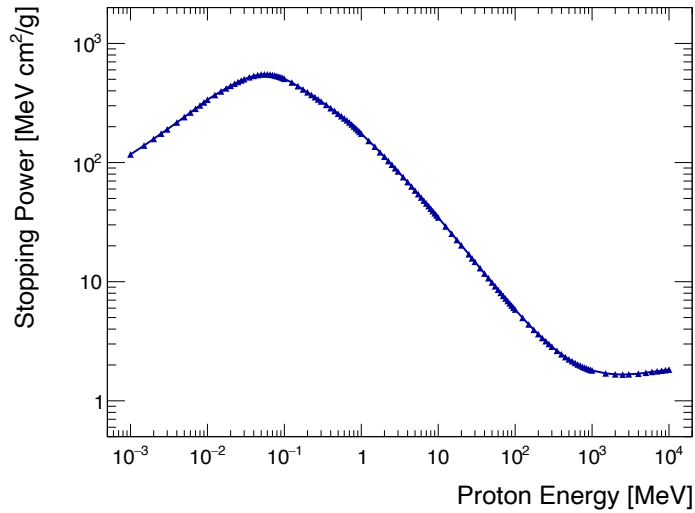

Figure 1. Stopping power for protons applied to silicon, from [14].

Applying Eq. 1 to each point of Figure 1, for $D_p$ = 0.189 mGy/day, yields Figure 2, proton fluence applied per day as a function of proton kinetic energy.

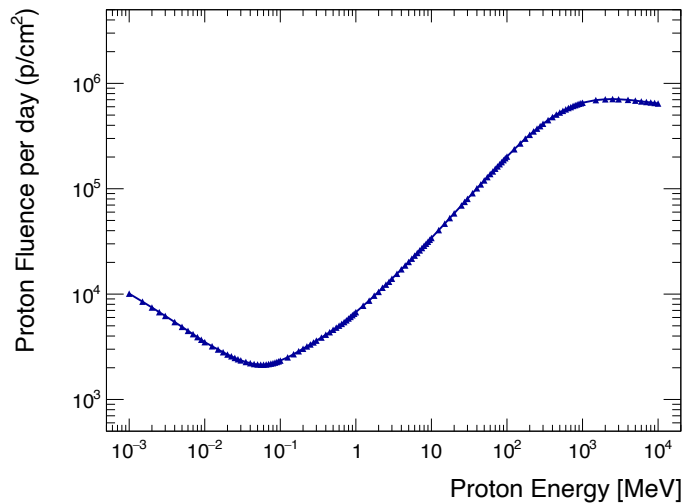

Figure 2. Proton fluence applied per day as a function of proton kinetic energy.

The information in Figure 2 can be converted from protons/cm$^2$ to n$_{eq}$/cm$^2$ using the NIEL scale factors, taken from Ref. [15], at each proton kinetic energy. The outcome is shown in Figure 3, the conversion between proton energy and fluence $\Phi_{eq}$, in n$_{eq}$/cm$^2$, per day.

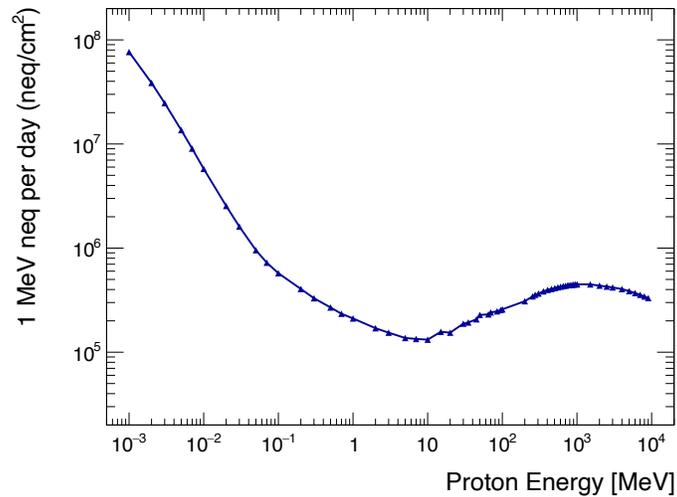

Figure 3. Average daily proton fluence, in 1-MeV-neutron equivalent per area, as a function of proton kinetic energy.

The energy spectrum to which silicon detectors operating on Mars will be exposed depends on their shielding. In the absence of knowledge of what that shielding (and hence spectrum) will be, we use Figure 3 to identify the worst-case baseline proton fluence scenario for the region of the Martian surface explored by the Curiosity Rover. Neglecting protons of kinetic energies below 1 MeV (as it is assumed [16] that these will not penetrate to the experimental apparatus), the worst-case maximum fluence rate would be observed for a spectrum dominated by protons of kinetic energy 1.005 GeV, the local maximum of Figure 3. The associated proton fluence rate is $4.49 \times 10^5$ $n_{eq}$/cm$^2$/day or 5.2 $n_{eq}$/cm$^2$/sec. This is rounded up to 6 $n_{eq}$/cm$^2$/sec and applied to the Hamburg Model simulation at 60 second intervals for a duration of 28.2 Earth-years to predict an integrated fluence of $5.3 \times 10^9$ $n_{eq}$/cm$^2$ for chronic solar protons and galactic cosmic ray protons of energies greater than or equal to 1 MeV.

A similar calculation is used to estimate the chronic fluence during transit, with the modification that the surface dose rate of 0.210 mGy/day is replaced by the measured [17] value for transit conditions, (higher in transit due to absence of protection by the Mars magnetosphere) of 0.481 mGy/day. The simulation applies this dose rate for 250 days, a period consistent with the length of time that was used to transport the Curiosity Rover to Mars.

In addition to the ambient flux of solar protons and galactic cosmic rays, silicon detectors at the Mars surface and in transit will be subjected to solar energetic particle (SEP) events. Two classes of SEPs have been reported [18], as categorized by their duration: the impulsive (1-20 hours) class, and the gradual (1-3 days) class. In this simulation we do not distinguish them: SEPs in this simulation are applied to the silicon for a duration of one Mars-day, 24 hours, 40 minutes.

The Curiosity Rover detected 1 SEP event in the first 300 sols (each of duration 24 hours 40 minutes) at the Martian surface [10]. On this basis, the simulation of conditions at the surface includes 2 SEPs per Mars-year, which is 30 SEPs in total. At the Mars surface, the observed [10] SEP dose is 0.025 mGy per event, which is $5.94 \times 10^4$ $n_{eq}$/cm$^2$/event. This is applied in the case of 27 of the simulated SEPs. The authors of Ref. [19] have noted, however, that "extreme" SEPs, which occur approximately once per solar cycle, should be considered separately due to their exceptional potential for damage. To incorporate extreme SEP radiation, we apply 3 such events over the 15 Mars-year duration. Each is applied for the duration of one Mars-day. Simulation of their effect uses a flux for them of $1.88 \times 10^{10}$ p/cm$^2$, corresponding to the maximum in the range reported in Ref. [19] for protons with energy greater than 30 MeV, equivalently $4.98 \times 10^5$ n$_{eq}$/cm$^2$/sec. Converted from protons to 1-MeV-neutron equivalent fluence, this predicts $4.41 \times 10^{10}$ n$_{eq}$/cm$^2$/event for each of these three cases. In the simulation, SEPs are applied to the Martian surface once in each Martian summer and once in each Martian winter. The extreme SEPS are applied during the Martian summer 2, 8, and 14 Mars-years after the sensor has reached the Mars surface, so there is one year of chronic fluence before the first extreme SEP and one year of chronic fluence after the third one.

The Curiosity Rover detected 5 SEPs during the 250 day transit [17]. Accordingly, to simulate a high impact scenario of SEP effects during the transit, we assume a total of 5 events, each with a dose (from Table 2 in Ref. [10]) of 19.5 mGy. A similar calculation to the one described above for computing chronic fluence on the surface is used to convert this dose rate to a fluence rate, corresponding to $4.63 \times 10^7$ n$_{eq}$/cm$^2$/event, or 536 n$_{eq}$/cm$^2$/sec.

This is sufficient as a fluence input to the Hamburg Model. The other inputs to the model include the depletion depth of the silicon sensor, the temperature, and the duration of exposure.

Temperatures of the Martian soil and atmosphere have been sampled at a variety of locations. Data reported [20] by the Rover Environmental Monitoring Station indicate an approximately sinusoidal variation in ambient air temperature over the course of one sol with the minimum at about 205 K and the maximum at about 270 K. Information from the Mars Pathfinder instrument indicates that surface temperature within 0.25 m of the ground can exceed the air temperature 1.5 m above the ground by 20 Kelvin degrees. Earlier measurements by Viking instruments estimated maximum summer surface temperatures at the equator to be about 293 K. During the winter, the temperature near the Martian poles diminishes to 248 K. To cover the worst case scenario, the simulation reported here uses ground temperature that varies sinusoidally with amplitude 40 K over the course of each sol. This cycle is modulated for seasonal variations during the Mars year: the mean temperature is set to 235.5 K with annual sinusoidal variation of amplitude 22.5 K. Thus the Martian-yearly maximum temperature is taken to be 298 K, and the Martian-yearly minimum is 173 K. This thermal profile is applied for 15 Martian years (28.2 Earth-years). The temperature during transit is conservatively taken to be 283 K, for consistency with the maximum temperature recorded [21] for the silicon detectors in the AMS experiment. AMS is used as a concrete example of a silicon detector functioning in the extra-terrestrial environment, whose thermal profile and operating characteristics are available in the literature. The maximum AMS temperature is used because leakage current and annealing rate both depend upon temperature, and the maximum permits a conservatively high prediction of leakage current.

The simulation examines the characteristics of sensors processed with two technologies: One uses the ATLAS design [22], based on oxygenated n-type bulk of depleted thickness 250 μm (representative of planar processed sensors). The other uses p-type bulk and sensor depletion distance 40 μm; this is taken to be approximately representative of 3D processed sensors [23], although the cylindrical shape of the electrodes, and correspondingly convergent pattern of the electric fields lines reaching them, is neglected. The initial donor concentration for the n-type sensors is taken to be $1.7 \times 10^{12}$ cm$^{-3}$. The initial acceptor concentration for the p-type sensors is taken to be $2 \times 10^{12}$ cm$^{-3}$. Both of these are typical values that are currently used in fabrications.

The change in leakage current $\Delta I$, in a sensor of active volume $v$, due to irradiation, is dependent on the fluence $\Phi_{eq}$ as $\Delta I/v = \alpha \Phi_{eq}$, where the damage constant, $\alpha$, describes both short- and long-term annealing [7]. The depletion voltage dependence on the effective doping concentration follows from the Poisson equation and is given as $V_{dep} = \dfrac{q|N_{eff}|d^2}{2\varepsilon\varepsilon_0}$, where $q$ is the charge of the primary carriers, $d$ is the depletion depth, $N_{eff}$ is the effective doping concentration, and $\varepsilon$ and $\varepsilon_0$ are the dielectric and vacuum permittivities, respectively.

The leakage current model [9] reflects several fluence-related effects upon the change, $\Delta N_{eff}$, in the doping concentration, which influences depletion voltage. These are discussed in detail in [8] and implemented in the n-type study here as in [24]. Modifications to the model, to reflect effects in p-type bulk, are described in [25]. The effects are associated with the rate of introduction of stable acceptors ($g_C\Phi_{eq}$, see Eq. 5.8 of [7]), whose parameter, $g_C$, is set to $0.43 \times 10^{-2}$ cm$^{-1}$. They also include the rate of short-term ("beneficial") annealing of acceptors $\Phi_{eq}\sum_i g_{a,i}\exp\left(-\dfrac{t}{\tau_{a,i}}\right)$, see Eq. 5.9 of Ref. [7]. The introduction rates are taken to be the same for all temperatures and one minute time intervals (the duration over which the irradiations are applied); this simulation uses an average introduction rate $g_a = 0.6 \times 10^{-2}$ cm$^{-1}$. A third contribution is the rate of anti-annealing ($g_Y\Phi_{eq}$, see Sect. 5.3.3 of Ref. [7]) whose parameter $g_Y = 6.0\times 10^{-2}$ cm$^{-1}$. These introduction rate values are chosen to match the values used for the B-Layer of the ATLAS Pixel detector [26]. The introduction rates reflect the calibrations in Figure 8 of Ref. [26]. Complete initial donor or acceptor removal is assumed to be possible. For the n-type simulation, the donor removal constant is $c^{donor} = 6.4 \times 10^{-14}$ cm$^2$, and for the p-type simulation, the acceptor removal constant is $c^{acceptor} = 1.98 \times 10^{-13}$ cm$^2$. For the p-type simulation, the short term annealing is taken to be negligible.

## 2    Simulation Results

Figure 4 (left) shows the change in leakage current density as a function of fluence in a sensor whose response is normalized to that at 294 K. Figure 4 (right) shows the same change in leakage current density, but as a function of the time, in Earth-years, from the beginning of the mission. The leakage current is clearly dominated by the extreme SEPs. The vertical drops represent annealing that dominates after the cessation of each extreme SEP. The curvature of the function in Figure 4 (left) in the fluence regime during application of each extreme SEP reflects the silicon's response to the temperature cycle over the course of a single Martian day. The step-like structure in Figure 4 (right) shows the response to the periodic non-extreme SEPs. The final value of $\Delta I_{leak}$, after 15 Mars-years, is 4.03 µA/cm$^3$. The peak leakage current, immediately after application of the final extreme SEP event, is 6.26 µA/cm$^3$.

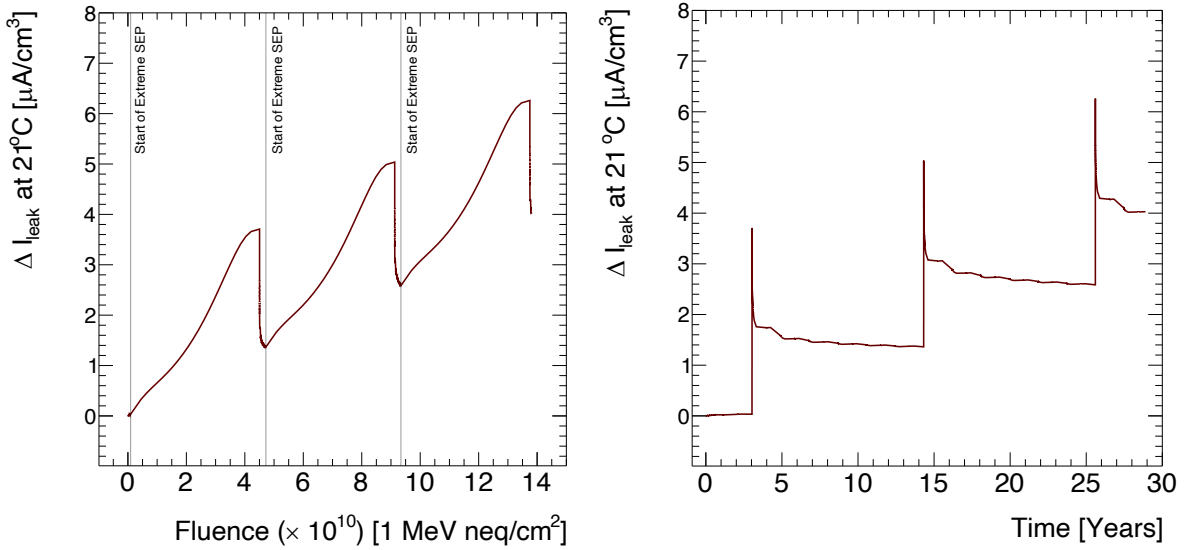

Figure 4.  Predicted change in leakage current density as a function of (left) integrated charged hadron fluence and (right) time in Earth-years, in both the n-type and p-type silicon sensors, normalized to 294 K.

To predict the response of silicon devices being operated in other Martian thermal environments $T$, the current at temperature $T_R = 294$ K can be translated according to [6]:

$$I_{leak}(T_R) = I_{leak}(T) \left( \frac{T_R}{T} \right)^2 \exp \left[ -\frac{E_{eff}}{2k_B} \left( \frac{1}{T_R} - \frac{1}{T} \right) \right],$$

(2)

where $E_{eff}$ is the effective silicon bandgap (approximately 1.21 eV) and $k_B$ is the Boltzmann constant.

Figure 5 shows the change in depletion voltage during the same interval, for a device with depletion depth 250 μm and oxygenated n-type bulk.  Figure 5 (left) shows the predicted full depletion voltage as a function of fluence, while Figure 5 (right) shows the same information versus time in Earth-years. The discontinuities reflect the effects of reverse annealing after cessation of extreme SEPs. The final value is 67 V.  The negative slope on the n-type device indicates that the sensors will not have undergone type inversion at this stage.  After inversion, the depletion voltage will rise; sensors of these geometries can continue to operate fully depleted up to several hundred volts.  Figure 6 shows depletion voltage versus (left) fluence and (right) time for a p-type sensor of depletion depth 40 μm, representative of 3D.  In this case the final value is approximately 2.4 V.  In the case of this p-type sensor, once the initial acceptor removal is exhausted, the effective acceptor introduction term will dominate subsequent behavior, and the slope will become positive [27].

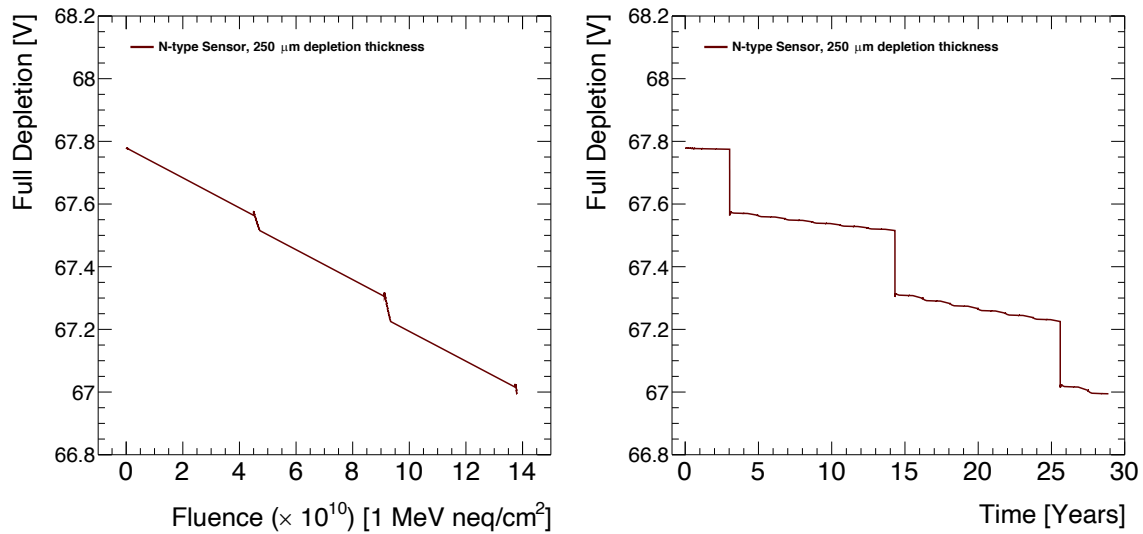

Figure 5. Depletion voltage as a function of (left) integrated charged particle fluence, and (right) time in Earth-years, for an oxygenated n-type silicon sensor of depletion depth 250 μm.

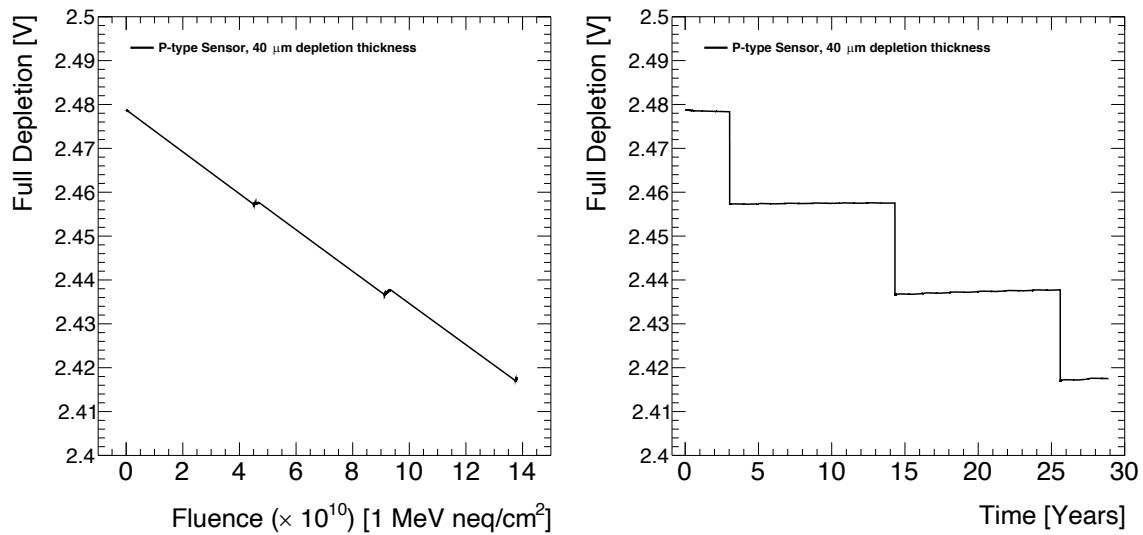

Figure 6. Depletion voltage as a function of (left) integrated charged particle fluence, and (right) time in Earth-years, for an p-type silicon sensor of depletion depth 40 μm.

## 3    Results, Uncertainties, and Discussion

In comparisons of silicon sensor leakage current data to predictions by this model, the model was found to under-predict the observed current by 15% to 47% [8] depending on the detailed conditions of the sensor. Accordingly the leakage currents predicted for silicon sensors in the Mars environment are conservatively scaled by a safety factor of 1.5. Including this safety factor, the predicted leakage current density and depletion voltages after 28.2 years are shown in Table 1 for the two sensor geometries considered. The current per volume is the most natural unit for indicating the leakage, because devices such as those considered here are operated in reverse-bias mode, in which the leakage current scales directly with volume; additionally the signal generated in the depleted region scales with depth. As a point of reference, the volume of a planar ATLAS pixel sensor (to which the data of Ref. [8] apply) is 0.25 cm$^3$.

The uncertainty on the predicted leakage current density is 10.5%, corresponding to contributions from the scale factor fit and the precision by which the temperature is known [8]. The contributions of uncertainties on the Hamburg Model parameters - approximately 20% - are not shown in the leakage current prediction. The uncertainty on the depletion voltage, due to uncertainties on the Hamburg Model parameters, is approximately 20%.

| Sensor Type | n-type (planar) | p-type (approximation to 3D) |
|---|---|---|
| Depletion Depth ($\mu$m) | 250 | 40 |
| Leakage Current density ($\mu$A/cm$^3$) | 9.4±1.0 | 9.4±1.0 |
| Depletion Voltage (V) | 67±13 | 2.4±0.5 |

Table 1. Predicted peak leakage current density and final depletion voltage for silicon sensors operating in the Martian environment for a duration of 28.2 Earth-years, including a safety factor of 1.5 applied to current. The 10.5% uncertainty on the leakage current corresponds to contributions from the scale factor fit and the precision by which the temperature is known. The uncertainty on the depletion voltage, due to uncertainties on the Hamburg Model parameters, is taken to be 20%.

The relevant fluence due to space radiation conditions over the duration of a rover mission to Mars is significantly lower than the fluence received by the silicon detectors in ATLAS [8,24]. Consistent with expected lower fluences, the leakage current predictions presented here are lower than in ATLAS. Similarly, the magnitude of the change in depletion voltage in planar sensors is smaller than in the ATLAS simulations. Depletion voltage predictions for 3D sensors in ATLAS have not been published, so no comparison may be made at this time to them. (Measurements of leakage current evolution in p-type bulk sensors are available in Ref. [24].). The simulation results presented here reflect the same expected annealing responses to temperature changes and received particle fluence as those in ATLAS simulations.

This work is restricted to the study of bulk silicon damage. Effects of surface damage by ionizing particles have not been included in the simulations reported here. The main surface effect is the increase of positive charge trapped in the surface oxide layer, which in turn can cause loss of isolation between electrodes, increased parasitic capacitance between adjacent regions, and variation in the distribution of the electric field at the Si-SiO$_2$ interface, affecting the breakdown voltage. Some of this can be mitigated by, for example, doping design in the inter-electrode region. A rough calculation predicts that these electronic devices in a space environment will be exposed to a gamma dose on the order of tens of krad from chronic sources. This modifies the above result by less than 1%, an amount that is unobservable within the uncertainty of the prediction. Gammas induced by SEPs contribute negligible additional dose [28]. The present authors have studied the effect of gamma irradiation on 3D silicon sensor operating current [29]; exposure to a 5 Mrad dose increased the operating current by approximately a factor of two.

The Perugia Model [30,31] incorporates surface effects, and it will provide an important foundation for follow-up studies to this one. Effective annealing behavior can be adapted from the Hamburg Model to the Perugia Model as is done in Ref. [24].

In the resources-limited environment of space, decreased power consumption can be realized with the 3D sensor technology. When possible, silicon sensors used for particle detection are typically operated above the estimated full depletion voltage to achieve optimal charge collection efficiency. In the case of 3D, these predictions indicate that an applied bias voltage as low as 10V may be sufficient for optimal operation.

In conclusion, Hamburg Model simulation of bulk electrical characteristics of silicon sensors using contemporary 3D and planar processing technology, fabricated with high resistivity n-type and p-type bulk, predicts that these devices' leakage current and depletion voltage will not pose restrictions for operation in Martian radiation and thermal conditions, including conditions associated with extreme Solar Energetic Particle events, through transit and for 28.2 Earth-years. These simulations may be used for insight and interpretation of measurements made with the Perseverance Rover [32]. The data collected by Perseverance may provide useful input and cross-checks of these simulations in the future.

## 4    Acknowledgments

This work was supported by the National Aeronautics and Space Administration (NASA) under Federal Award Numbers NNX15AL51H (2018-RIG) and 80NSSC20M0034 (2020-RIG), and by the U.S. Department of Energy award DE-SC0020255. The support of Dr. Paulo Oemig of New Mexico State University and the encouragement of Dr. Insoo Jun of Jet Propulsion Laboratory/NASA are deeply appreciated.